\documentstyle[twocolumn,aps,prl,psfig,floats]{revtex}

\begin{document}
\title{Nuclear Saturation with QCD sum rules}
\author{L. A. Barreiro}
\address{Instituto de Geoci\^encias e Ci\^encias Exatas, \\
Departamento de F\'{\i}sica - UNESP, \\
Cx.P. 178, CEP 13500-970, Rio Claro, SP, Brazil}

\address{\parbox{14.5cm}{\rm\small
\medskip\bigskip
\centerline{\bf Abstract}
The scalar and vectorial self energies obtained through QCD sum rules are
introduced in the Quantum Hadrodynamics (QHD) equations. This QHD and QCD
mixing show us that the effect of the density on the coupling constants is
very small.
}}
\maketitle

\section{Introduction}

The research for the exact equilibrium properties of the nuclear matter \cite
{Wal,Chin} is an old problem of nuclear physics and it continues to have a
great interest nowadays. Quantum Hadrodynamics (QHD) \cite{SeW} is one of
the models with larger success to describe the properties of the infinite
nuclear matter as well as those of finite nuclei. QHD describes the nucleon-nucleon
interaction through the mesons exchange ($\pi $, $\sigma $, $\omega $, $\rho 
$, etc.). Several calculations of nuclear structure using QHD and its
extensions were made with success in the explanation of experimental data 
\cite{LSR}. For the purposes of this work only the
simplest QHD model will be considered.

The basic QHD-model that explains the nuclear matter includes the nucleon ($%
\psi $) coupled with sigma ($\sigma $) and omega\ ($\omega $) mesons. In
spite of the pion be the principal component of the nucleon-nucleon
interaction, it is not included because the nuclear matter is an isotropic
system with parity conservation. Analyzing the model we see that the real
part of the scalar term ($\sigma $-meson) is typically of the order of
several hundred MeV attractive while the real part of\ the time component of
the vector term ($\omega $-meson) is typically of the order of several
hundred MeV repulsive. However the energies involved in problems of nuclear
structure are only of a few tens of MeV. That order of energy is obtained on
the QHD model due to a large cancellation among the scalar and vector pieces
and this process of cancelation can be controlled through an appropriate
choice of the coupling constants. However, these so called QHD constants are
quite different from those ones given in Bonn potential \cite{Bonn} or
empirical data \cite{SeW}. So it is interesting to discuss the validity of
these coupling constants.

The success of the Walecka model, which is based on Dirac's equation, is
due to the mutual cancellation between the large scalar and vector
potentials. However in some works \cite{Brody,BBJ} it is discussed that the
composed nature of the nucleon can suppress the scalar optical potential.
However, as it is argued in \cite{CFGX}, it can be shown \cite{WGT} that
due to the nucleon being immersed in the nuclear media, such suppression
does not exist. This result makes possible the use of the Dirac
phenomenology for composed particles. Besides, recently the effective field
theory (EFT) \cite{Wein1,Wein2,kaplan,Furnst1,Furnst2,SeW2} has verified that 
the QHD models are
consistent with the symmetries of the Quantum Chromodynamics (QCD), the
correct theory of the strong interaction. This last fact motivates the mixing
between QHD and QCD accomplished in this work.

It is known that QCD is the correct theory of the strong interaction.
However, there is not a perturbative treatment for QCD at energies involved
in nuclear matter problems, so it is interesting to use the nonperturbative
method given by QCD sum rules that was introduced by Shifman, Vainshtein,
and Zakharov in the late 1970's \cite{SVZ}. The method consists in describing
a correlation function in terms of hadron degree of freedom as well as
quarks degree of freedom. This last one can be written in an operator
product expansion (OPE) where the perturbative part (Wilson coefficients) is
separate from the nonperturbative (condensates). The power of this technique
is in the fact that the nonperturbative part is the same for all problems.

The objective of this work is to get the results obtained from QCD sum rules
in the media \cite{CFGX} and use them on QHD equations to analyze the
necessary constants to obtain the saturation point of the infinite nuclear
matter.

\section{The QHD Model}

The model used to describe the nuclear matter is from Serot and Walecka 
\cite{SeW} and it includes nucleons ($\psi $) interacting with $\sigma $ and 
$\omega $ mesons. So that the Lagrangian density is given by 
\begin{eqnarray}
{\cal L} &=&\bar{\psi}(\gamma ^{\mu }\partial _{\mu }+g_{s}\sigma
-g_{v}\gamma ^{\mu }\omega _{\mu }-M)\psi +  \nonumber \\
&&+\frac{1}{2}(\partial _{\mu }\sigma \partial ^{\mu }\sigma
-m_{s}^{2}\sigma ^{2})-\frac{1}{2}\left( G_{\mu \nu }G^{\mu \nu
}-m_{v}^{2}\omega ^{\mu }\omega _{\mu }\right) \;,  \label{lagrangian}
\end{eqnarray}
where $g_{s}$ and $g_{v}$ are the mesons coupling constants, $G_{\mu \nu
}=\partial _{\mu }\omega _{\nu }-\partial _{\nu }\omega _{\mu }$ is the
strength tensor for the vector field and $M$, $m_{s}$ and $m_{v}$ are the
nucleon, $\sigma $-meson and $\omega $-meson masses, respectively. The Green
function for the nucleon is given by 
\begin{eqnarray}
G(k)&=&\left( \gamma _{\mu }\tilde{k}^{\mu }-M^{\ast }\right) \left\{ \frac{1}{%
\tilde{k}^{2}-M^{\ast 2}+i\epsilon }+\right. \nonumber \\
&& \left. + \frac{i\pi }{E^{\ast }(k)}\delta \left(
k^{0}-E(k)\right) \theta (k_{F}-|{\bf k|})\right\} \;,  \label{nucleon}
\end{eqnarray}
where it was defined 
\begin{eqnarray}
\tilde{k}^{\mu } &\equiv &k^{\mu }+\Sigma _{(v)}^{\mu }\;, \\
M^{\ast } &\equiv &M+\Sigma _{(s)}\;,  \label{mas_star} \\
E^{\ast }(k) &\equiv &\sqrt{|{\bf \tilde{k}|}^{2}+M^{\ast 2}}\;.
\end{eqnarray}
Here $\Sigma _{(s)}$ and $\Sigma _{(v)}^{\mu }$ represent the scalar and
vector self-energies, respectively.

In the more simple model only the contributions of the baryons from the Fermi 
sea will be considered. This is equivalent to disregard the contributions 
coming from antibarions (Dirac sea). So in agrement with the definition of
energy-momentum tensor 
\begin{eqnarray}
\varepsilon  &=&\langle \Psi |\hat{T}_{00}|\Psi \rangle -VEV  \nonumber \\
&=&\frac{1}{2}\frac{m_{s}^{2}}{g_{s}^{2}}\Sigma _{(s)}^{2}+\frac{1}{2}\frac{%
m_{v}^{2}}{g_{v}^{2}}\Sigma _{(v)}^{2}+\frac{\gamma }{(2\pi )^{3}}%
\int_{0}^{k_{F}}\text{d}^{3}k\sqrt{|{\bf k|}^{2}+M^{\ast 2}}\;,  \nonumber \\
&&  \label{e-mft}
\end{eqnarray}
where $VEV$ is the vacuum expectation value of the $\hat{T}_{00}$ and $%
\gamma $ is the spin-isospin degeneracy. The self-energies can be obtained
through ``tadpole'' Feynman diagrams and are given by the following
relationships \ 
\begin{eqnarray}
\Sigma _{(s)} &=&M^{\ast }-M  \nonumber \\
&=&-\frac{g_{s}^{2}}{m_{s}^{2}}\frac{\gamma M^{\ast }}{4\pi ^{2}}\left[
k_{F}E_{F}^{\ast }-M^{\ast 2}\ln \left( \frac{k_{F}+E_{F}^{\ast }}{M^{\ast }}%
\right) \right]   \label{sigma_s}
\end{eqnarray}
and 
\begin{equation}
\Sigma _{(v)}^{\mu }=-\delta ^{\mu 0}\frac{g_{v}^{2}}{m_{v}^{2}}\frac{\gamma 
}{6\pi ^{2}}k_{F}^{3}\;,
\end{equation}
The expression (\ref{sigma_s}) should be solved in a self-consistent way.
Therefore all the ingredients are available to calculate the energy density.
This result is also known as Mean Field Theory (MFT), because it can be
obtained with a mean field approach for the meson fields.

A simple analysis shows us that the self-energies $\Sigma _{(s)}$ and $%
\Sigma _{(v)}$ are proportional to the scalar and vectorial densities,
respectively. Then the form presented by Eq. (\ref{e-mft}) represents an
expansion on powers of these densities. On the other hand, in agreement with
the EFT concepts, the energy density is a functional and has an expansion
in powers of the densities (scalar, vectorial, etc.) organized through 
Georgi's naive dimensional analysis (NDA). So, the energy densities have the
same form that the expansions obtained by EFT\cite{Furnst1,Furnst2,SeW2}. Therefore,
since QHD has a foundation on the symmetries of QCD, the next step is to
calculate the self-energies through the QCD methods. However, as the energy
levels treated in the nuclear problems are in a nonperturbative regime for
QCD, the sum rules is the appropriate method to be used.

\section{QCD Sum Rules}

\label{sectx}

The QCD sum rules method \cite{SVZ} begins with the following time-ordered\
correlation function, defined by 
\begin{equation}
\Pi _{\alpha \beta }(q)\equiv i\int \text{d}^{4}x\text{ e}^{iq\cdot
x}\langle 0|T\left[ \eta _{\alpha }(x)\bar{\eta}_{\beta }(0)\right]
|0\rangle \;,  \label{correlator}
\end{equation}
where $|0\rangle $ is the physical nonperturbative vacuum state and $\eta
_{\alpha }(x)$ is an interpolating field with the quantum numbers of a
nucleon. As proposed by Ioffe \cite{IOF}, the proton field is given by 
\begin{equation}
\eta (x)=\epsilon _{abc}\left( u_{a}^{T}C\gamma _{\mu }u_{b}\right) \gamma
_{5}\gamma ^{\mu }d_{c}\;,
\end{equation}
where $u$ and $d$ are the quark fields, $a,$ $b,$ $c$ are the color indexes
and $C$ is the charge-conjugation matrix. Now the correlation function can
be written as an operator product expansion (OPE) whose nonperturbative
part (condensates) can be separated from the perturbative one (Wilson
coefficients). The OPE can be generated starting from the following
expansion for the quark propagator\cite{RRY,KCY} 
\begin{eqnarray}
S_{ab}(x)&=&\langle 0|T\left[ q_{a}(x)\bar{q}_{b}(0)\right] |0\rangle \nonumber \\ 
&=&i\frac{\delta _{ab}}{2\pi ^{2}}\frac{\not{x}}{x^{4}}-\frac{\delta _{ab}}{12}\langle 
\bar{q}q\rangle _{vac}+\cdots \;,
\end{eqnarray}
where $\langle \bar{q}q\rangle _{vac}$ is the quark condensate which can be
determined from the Gell-Mann-Oakes-Renner relation, 
\[
2m_{q}\langle \bar{q}q\rangle _{vac}=-m_{\pi }^{2}f_{\pi }^{2}\left( 1+{\cal %
O}(m_{\pi }^{2})\right) \;. 
\]
In this relation, $m_{\pi }=138$ MeV is the pion mass, $f_{\pi }=93$ MeV is
the pion decay constant and $m_{q}=(m_{u}+m_{d})/2\simeq 7\pm 2$ MeV is the
average of the up and down quark masses.

The same correlation function, Eq (\ref{correlator}), can be written in a
phenomenological way, which mean that the correlator has a hadronic
description through the nucleon Green function, Eq (\ref{nucleon}). The
match of the theoretical side (OPE) and the phenomenological one is the
essence of the QCD sum rules. Using the fundamental state of the nuclear
matter as vacuum and the interacting propagator in\ the phenomenological
side, it is possible to obtain the following sum rules \cite{CFGX} 
\begin{eqnarray}
M_{N}^{\ast } &=&-\frac{8\pi ^{2}}{M^{2}}\langle \bar{q}q\rangle _{\rho
_{B}}\;,  \label{ms} \\
\Sigma _{(v-QCD)} &=&\frac{64\pi ^{2}}{3M^{2}}\langle q^{\dagger }q\rangle
_{\rho _{B}}\;.  \label{mv}
\end{eqnarray}
Here $M^{2}$ represent the borel mass with value near $M^{2}=1$GeV$^{2}$\cite
{IOF}, and 
\begin{eqnarray}
\langle q^{\dagger }q\rangle _{\rho _{B}} &=&\frac{3}{2}\rho _{B}\;, \\
\langle \bar{q}q\rangle _{\rho _{B}} &=&\left( 1-\frac{\sigma _{B}\rho _{B}}{%
m_{\pi }^{2}f_{\pi }^{2}}+\cdots \right) \langle \bar{q}q\rangle _{vac}\;.
\end{eqnarray}
The expression (\ref{ms}) is a generalization of the Ioffe's formula \cite
{IOF} to finite density. The sigma term is estimate in Ref \cite{GLS} as $%
\sigma _{B}\simeq 45\pm 10$ MeV. Taking the ratios of the expressions (\ref
{ms}) and (\ref{mv}) in relation to the mass $M_{N}$ (Ioffe's formula), they
are obtained the following results: 
\begin{eqnarray}
\Sigma _{(s-QCD)} &=&-\frac{\sigma _{B}M_{N}}{m_{\pi }^{2}f_{\pi }^{2}}\rho
_{B}\;,  \label{self_qcd_s} \\
\Sigma _{(v-QCD)} &=&\frac{8m_{q}M_{N}}{m_{\pi }^{2}f_{\pi }^{2}}\rho _{B}\;.
\label{self_qcd_v}
\end{eqnarray}
Although a self-consistent relation is not obtained as in Eq (\ref{sigma_s}%
), these expressions have a more fundamental nature based on quark and gluon
degree of freedom. At this point, the idea is to use the expressions (\ref
{self_qcd_s}) and (\ref{self_qcd_v}) to calculate the energy density of the
nuclear matter.

\section{Numerical Results and Conclusion}

\label{secty}

Firstly it should be noted that only the ratios between the coupling
constants and the mesons mass appear in the expressions where the energy
density of the nuclear matter is calculated. Some values for these ratios
are presented in the table \ref{tab1}.
 \begin{table}[b] 
\begin{tabular}{ccc} 
& $g_{s}^{2}/m_{s}^{2}$ & $g_{v}^{2}/m_{v}^{2}$ \\ \hline 
MFT & 3.029 10$^{-4}$ & 2.222 10$^{-4}$ \\  
Bonn Potential & 3.233 10$^{-4}$ & 4.104 10$^{-4}$ \\  
Empirical Values & 3.402 10$^{-4}$ & 3.516 10$^{-4}$%
\end{tabular} 
\caption{Coupling constants used in MFT and in Bonn potential and the 
empirical values. } 
\label{tab1} 
\end{table}   
In the first line of table \ref{tab1} the coupling constants were
adjusted so that the MFT might reproduce the saturation properties of the nuclear
matter. With the use of these coupling constants, the result obtained with
MFT are the following (dotted curve on Fig. 1): saturation point on $%
k_{F}=1.12$ fm$^{-1}$ with a energy density by nucleon given by $-15.75$
MeV. This result reproduces the exact equilibrium properties of the nuclear
matter given by Refs \cite{Wal,Chin}, which is the expected result, once the
coupling constants were chosen for such adjustment to occur. However, these
coupling constants are very different of the empirical values \cite{SeW}.

\begin{figure}[h] 
\centerline{\psfig{figure=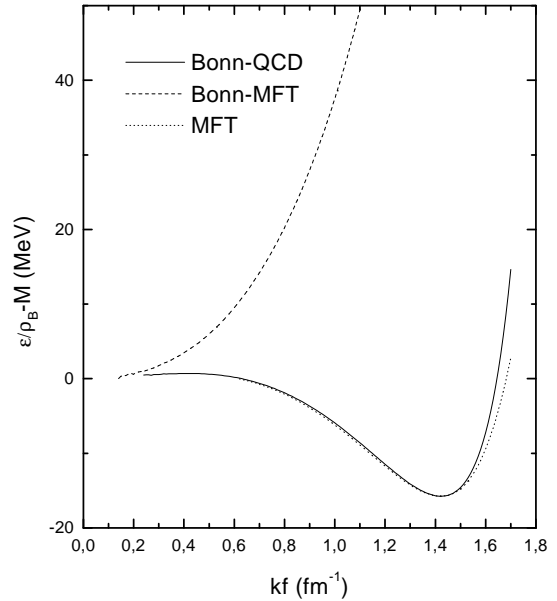,height=4.5in}} 
\caption{Binding energy by nucleon in nuclear matter as function of Fermi 
momentum. In the continuous line is adopted the Bonn coupling constants and the  
self-energies obtained with QCD sum rules. In the dotted and dashed lines, the original  
MFT equations are used together with the coupling constants from the MFT and Bonn  
potential, respectively. }  
\label{fig1} 
\end{figure} 

The most accepted values for these constants in free space are found in
the Bonn potential \cite{Bonn}, for which the nucleon-nucleon interaction is
adjusted to describe the scattering data. The Bonn coupling constants for
the sigma and omega mesons are given on the second line of table \ref
{tab1} while the empirical values are presented on the third line of that
table. It must be noted that, in MFT the value of the coupling
constant for the scalar meson is higher than the respective value for the
vector meson. But in the Bonn potential and in the empirical data there is
an inversion in the magnitude of these constants. Using the Bonn's values in
Eq. (\ref{e-mft}), the result presented by the dashed curve on Fig. 1  is 
obtained. That curve shows us that the energy density for MFT does not present a good
behavior. There is no saturation point, in other words, there is no
formation of nuclear matter. The same result is obtained with the empirical
constants. The problem is in the fact that the use of the Bonn or empirical
coupling constants implicates in a small increase of the attraction but with
a much larger increase in the repulsion. It can be argued that the
difference between the QHD coupling constants and the Bonn values is due to
the fact that the latter are obtained in free space while the QHD
values are obtained in nuclear media, but this argument is not valid for
empirical constants. However, using the constants given by Bonn potential,
and the self-energies found through QCD sum rules, Eqs. (\ref{self_qcd_s})
and (\ref{self_qcd_v}), with $\sigma _{B}=49$ MeV and $m_{q}=6.8$ MeV, the 
result represented by the continuous line on Fig. 1 is obtained. 
That curve
has a saturation point $k_{F}=1.42$ fm$^{-1}$ with $-15.75$ MeV for the
energy density by nucleon, which is the saturation point of the nuclear
matter. Therefore the QCD sum rules allows us to reproduce the properties of
nuclear matter with the simplest QHD model using more realistic values for
the coupling constants.

Since the empiric constants are not so different from the Bonn values when
they are compared to the QHD values, it can be that the effects of the
density do not alter the values of these constants significantly. On the other hand,
the exact determination of the parameters $\sigma _{B}$ and $m_{q}$ should
allow a better evaluation of the coupling constants in the media for the
various QHD models and consequently to determine with more precision their
real variation in relation to the values of the constants in the vacuum.
Furthermore this result is another indicative that the quark degree of
freedom takes an important place in nuclear problems.

Finally, it is known that the QCD sum rules work\ with more accuracy when
there is high transferred momentum. So I hope this mix between QCD sum rules
and QHD model can be applied, with success, in systems in a regime of high
density and high temperature as neutron stars. Besides, in these calculations 
the contributions of the gluon condensate and four quark condensate were not 
included. But it is known that the contribution of these terms for the sum
rule is very small. However, if more precise answers are required, these
terms have to be included as well as the most sophisticated versions of QHD.
These and other questions are left for future works.

\begin{center}
\noindent {\bf Acknowledgments}
\end{center}

\noindent I would like to thank Prof. G. Krein for the discussions and
suggestions and also FAPESP (Funda\c{c}\~{a}o de Amparo \`{a} Pesquisa do
Estado de S\~{a}o Paulo) for the financial support.


\begin{references}
\bibitem{Wal}  J. D. Walecka, Phys. Lett. {\bf B 59} (1975) 109.

\bibitem{Chin}  S. A. Chin, Phys. Lett. {\bf B 62} (1976) 263.

\bibitem{SeW}  B. D. Serot and J. D. Walecka, Adv. Nuc. Phys., {\bf 16}
(1986) 1.

\bibitem{LSR}  G. A. Lalazissis, M. M. Sharma, and P. Ring, Nucl. Phys. {\bf %
A 597} (1996) 35.

\bibitem{Bonn}  R. Machleidt, K. Holinde, and Ch. Elster, Phys. Rep. {\bf 149%
} (1987) 1.

\bibitem{Brody}  S. J. Brodsky, Comm. Nucl. Part. Phys. {\bf 12} (1984) 213.

\bibitem{BBJ}  E. Bleszynski, M. Bleszynski, and T. Jaroszewicz, Phys. Rev.
Lett. {\bf 59} (1987) 423.

\bibitem{CFGX}  T. D. Cohen, R. J. Furnstahl, D. K. Griegel, and Xuemin Jin,
Prog. Part. Nucl. Phys. {\bf 35} (1995) 221.

\bibitem{WGT}  S. J. Wallace, F. Gross, and J. A. Tijon, University of
Maryland Report No. 95-020 (1994).

\bibitem{Wein1}  S. Weinberg, Physica {\bf A 96} (1979) 327.

\bibitem{Wein2}  S. Weinberg, The Quantum Theory of Fields, vol {\bf I}:
Foundations (Cambridge University Press, New York, 1995).

\bibitem{kaplan}  D. Kaplan, Effective Field Theories, from: lectures given
as Seventh Sommer School in Nuclear Physics at the Institute for Nuclear
Theory, June 1995.

\bibitem{Furnst1}  J. J. Rusnak and R. J. Furnstahl, Nucl. Phys. {\bf A 627}
(1997) 495.

\bibitem{Furnst2}  R. J. Furnstahl, B. D. Serot and H. -B. Tang, Nucl. Phys. 
{\bf A 615} (1997) 441.

\bibitem{SeW2}  B. D. Serot and J. D. Walecka, Int. J. Mod. Phys., {\bf E 6}
(1997) 515.

\bibitem{SVZ}  M. A. Shifman, A. I. Vainshtein, and V. I. Zakharov, Nucl.
Phys. {\bf B 147} (1979) 385.

\bibitem{IOF}  B. L. Ioffe, Nucl. Phys. {\bf B 188} (1981) 317.

\bibitem{RRY}  L. J. Reinders, H. Rubinstein, and S. Yazaki, Phys. Rep. {\bf %
127} (1985) 1.

\bibitem{KCY}  K. C. Yang et al., Phys Rev {\bf D 47} (1993) 3001.

\bibitem{GLS}  J. Gasser, H. Leutwyler, and M. E. Sainio, Phys. Lett. {\bf B
253} (1991) 252.
\end{references}
\end{document}